\documentclass{revtex4}
\usepackage{amsmath,amsthm,amssymb,graphicx}

\def\beq#1\eeq{\begin{equation}#1\end{equation}}
\newcommand{\Schrodinger}{Schr\"odinger}
\newcommand{\Bayin}{Bay\i n}
\newcommand{\Riesz}{\left(-\hbar^2\triangle\right)^{\alpha/2}}
\newcommand{\abs}[1]{\lvert#1\rvert}
\newcommand{\C}{\mathcal C}

\newcommand{\cc}{\mathrm{c.c.}}

\begin{document}
\title{Comment on: ``On the consistency of solutions\\ of the space fractional \Schrodinger\ equation''}

\author{E. Hawkins$^1$\footnote{mrmuon@mac.com} and J.\ M.\ Schwarz$^2$\footnote{jschwarz@physics.syr.edu}}
\affiliation{$^1$Department of Mathematics, University of York, UK \\$^2$Physics Department,
Syracuse University,
Syracuse, NY, 13244, USA}
\begin{abstract}
In [\textit{J.\ Math.\ Phys.} \textbf{53}, 042105 (2012)], \Bayin\ claims to prove the consistency of the purported piece-wise solutions to the fractional \Schrodinger\ equation for an infinite square well. However, his calculation uses standard contour integral techniques despite the absence of an analytic integrand.  The correct calculation is presented and supports our earlier work proving that the purported piece-wise solutions do {\it not} solve the fractional \Schrodinger\ equation for an infinite square well [M. Jeng, S.-L.-Y. Xu, E. Hawkins, and J. M. Schwarz, {\it J. Math. Phys.} {\bf 51}, 062102 (2010)].  
\end{abstract}

\maketitle
The one-dimensional fractional \Schrodinger\ equation \cite{Laskin2000,Laskin2002} is given by
\begin{equation}
\label{fractional}
i\hbar \frac{\partial\psi (x,t)}{\partial t} =
D_\alpha \Riesz \psi (x,t)+
V(x,t)\, \psi (x,t),
\end{equation}
where $D_\alpha$ is a constant, 
$\Delta\equiv \partial^2/\partial x^2$ is the Laplacian, 
and $\Riesz$ is the quantum Riesz
fractional derivative, which is defined as 
\begin{equation}
\Riesz \psi(x,t) \equiv
\frac{1}{2\pi\hbar} \int_{-\infty}^{+\infty} 
e^{ipx/\hbar} \left| p \right|^\alpha \phi(p,t)\, dp\ , 
\end{equation}
where $\phi(p, t)$ is the Fourier transform of the
wavefunction,  
\begin{equation}
\phi(p,t) \equiv \int_{-\infty}^{+\infty} \psi(x,t)\,
e^{-ipx/\hbar}\, dx\ .
\end{equation}
When $\alpha=2$, the quantum Riesz fractional
derivative becomes equivalent to the ordinary Laplacian, and
we recover the ordinary Schr\"{o}dinger equation. 

Unless $\alpha$ is an even natural number, the quantum Riesz fractional derivative is a nonlocal operator, and  the usual piece-wise approach to solving the ordinary \Schrodinger\ equation is inapplicable. Despite this, in Ref.~\cite{Laskin2000}, Laskin used the piece-wise approach  for the fractional \Schrodinger\ equation with an infinite square well potential; in particular, he claimed that
\beq
\psi_0(x) = 
\begin{cases}
A \cos \left(\frac{\pi x}{2a}\right) & 
\mathrm{for}\ \abs{x}\leq a \\
0 & \mathrm{otherwise},
\end{cases}
\label{ground state}
\eeq
is the ground state solution. In Ref.~\cite{us}, we demonstrated that for $-1<\alpha<1$ and $\alpha\neq 0$, this $\psi_0(x)$ is not a solution. If $\psi_0(x)$ were a solution, then $\Riesz \psi_0(x)$ would vanish at $x=a$ (one of the ends of the infinite square well) but we showed that it does not. More specifically, $\Riesz\psi_0(a)$ is proportional to
\begin{equation}
f(\alpha)\equiv \int_0^\infty \frac{p^\alpha}{p^2-1}
\cos^2\left(\frac{1}{2}\pi p\right) dp\ ,
\end{equation}
and we showed that $df/d\alpha>0$, therefore $f(\alpha)$ only vanishes for $\alpha=0$.  The integral does not need to be explicitly evaluated to reach this conclusion.

In Ref.~\cite{Bayin}, \Bayin\ claims to evaluate $\Riesz \psi_0(x)$ explicitly to (incorrectly) conclude that $f(\alpha)=0$ for all $\alpha$.  To clarify this matter further, we not only point out several errors in Ref.~\cite{Bayin}, but also provide the correct calculation.      

\section{Errors}
\Bayin\ begins with the fractional \Schrodinger\ equation \eqref{fractional} and claims that with an infinite square well potential, this is equivalent, after separating variables,  to his Eq.~(6) --- the free, time-independent, fractional \Schrodinger\ equation with the ``boundary'' conditions $\psi(-a)=\psi(a)=0$.

These are not equivalent. Indeed, these are not even equivalent for the ordinary \Schrodinger\ equation. The solutions of the \Schrodinger\ equation with an infinite square well vanish outside $[-a,a]$, whereas the solutions of the free \Schrodinger\ equation with these boundary conditions are sine waves that continue outside of $[-a,a]$.

The problem is (at least partly) that \Bayin\ has ignored the point of our paper~\cite{us}. He claims in his conclusions that ``the solution inside the well is consistent with the outside.'' Because the fractional \Schrodinger\ equation is nonlocal, it is meaningless to speak of solutions inside and outside. The system must be treated as a unified whole.

More concretely, \Bayin\ considers the integral
\beq
\label{Bayin}
I = \int_{-\infty}^{+\infty} \frac{\abs{q}^\alpha \cos(\pi q/2)}{q^2-1}e^{i\pi q x/2a}\,dq \ ,
\eeq
where $q=\frac{2a}{\pi\hbar}p$, 
and claims that this is ``a singular integral with poles on the real axis at $q=\pm1$''. In fact, the integrand is continuous and bounded, because the zeroes of the cosine cancel the zeroes of the denominator.

The most serious error is his claim that \eqref{Bayin} can ``be evaluated via analytic continuation as a Cauchy principal value integral.'' 
After breaking the integral \eqref{Bayin} into pieces, he claims to evaluate these pieces by closing the contour in the upper or lower half plane. This technique requires an integrand that is an analytic function of $q$ over the relevant half plane.

Most of the factors in the integrand are explicitly analytic functions, but the trouble is the factor of $\abs{q}^\alpha$. This is equal to $q^\alpha$ for $q>0$ and $(-q)^\alpha$ for $q<0$. These functions have obvious analytic continuations, but it is impossible to join them together, because they are never equal for $q\neq0$, therefore this integrand does not have an analytic continuation to the upper or lower half plane. 

In Ref.~\cite{Bayin2}, \Bayin\ claims that in Ref.~\cite{Bayin} he analytically continued the integrand by the ``replacement'' of $\abs{q}^\alpha$ with
\beq
\label{replacement}
\frac{(iq)^\alpha+(-iq)^\alpha}{2\cos(\alpha\pi/2)} \ .
\eeq
This expression is ambiguous. It is only equal to $\abs{q}^\alpha$ for real $q$ if we take the branch cut for the first term in the upper half plane, but the branch cut for the second term in the lower half plane; however, that does not define an analytic function over either half plane. There are infinitely many possible interpretations of \eqref{replacement} --- corresponding to the possible choices of $i^\alpha$ and $(-i)^\alpha$ --- but none of them reproduces \Bayin's evaluation of $I$.

%

\section{Analytic computation}
\begin{figure}
\scalebox{.5}{\includegraphics{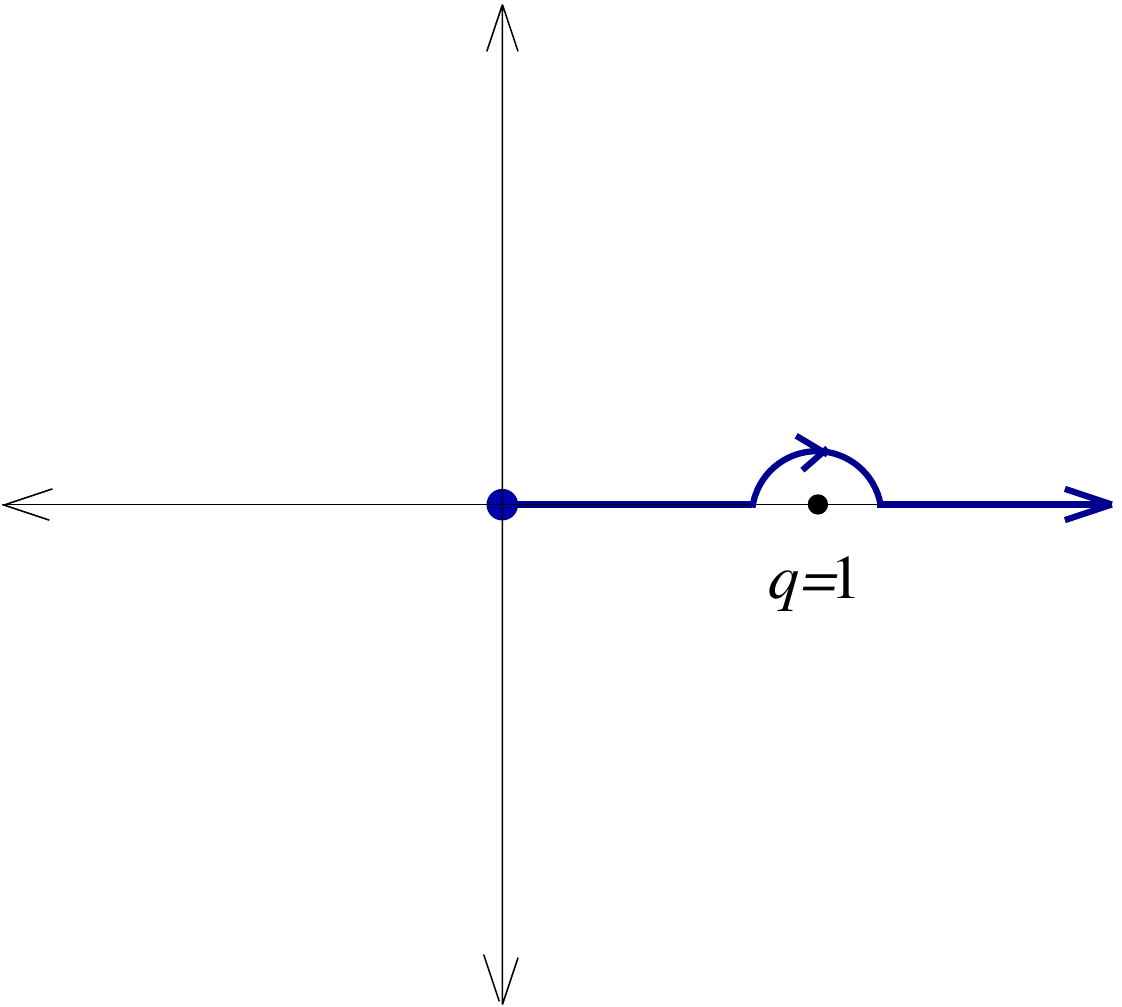}}
\caption{The contour, $\C$.\label{Contour}}
\end{figure}

It \emph{is} possible to evaluate the integral \eqref{Bayin} in closed form, but this is not as elementary as \Bayin\ claims.
First, note that the integral can be trivially rewritten as
\beq
\label{nicer form}
I(x) = 2\int_{0}^{+\infty} \frac{q^\alpha}{q^2-1}\cos(\pi q/2) \cos(\pi q x/2a) \,dq\ .
\eeq
This integrand is easily continued to an analytic function of $q$ by taking the branch cut of $q^\alpha$ along the negative real axis. With this, the integral \eqref{nicer form} can be equivalently taken along a contour, $\C$, that follows the positive real axis, except for a small clockwise detour around $q=1$; see Figure~\ref{Contour}.

The product of cosines can be written as a sum of 4 exponential functions. It is convenient to break the integral \eqref{nicer form} up in this way. Now, define
\beq
\label{J integral}
J(\lambda) \equiv \int_\C \frac{q^\alpha}{q^2-1}e^{i\lambda q} \,dq \ ,
\eeq
so that 
\[
I(x) = \tfrac1{2}\left[J\left(\tfrac\pi2\left[\tfrac{x}{a}+1\right]\right) + J\left(\tfrac\pi2\left[-\tfrac{x}{a}-1\right]\right) + J\left(\tfrac\pi2\left[-\tfrac{x}{a}+1\right]\right) + J\left(\tfrac\pi2\left[\tfrac{x}{a}-1\right]\right)\right] \ .
\]

In order to evaluate $J(\lambda)$, we will need the related integral
\beq
\label{K int}
K(\lambda) \equiv \int_0^\infty \frac{t^\alpha}{t^2+1}e^{-\lambda t} dt 
\eeq
for $\lambda\geq0$.
To compute this, first note that
\begin{align*}
K(0) &= \int_0^\infty \frac{t^\alpha}{t^2+1}dt
= \int_0^{\pi/2} \tan^\alpha \theta\,d\theta 
= \tfrac12 B(\tfrac12-\tfrac12 \alpha,\tfrac12 + \tfrac12\alpha) \\
&= \frac{\Gamma(\tfrac12-\tfrac12 \alpha)\Gamma(\tfrac12 + \tfrac12\alpha)}{2\,\Gamma(1)}
= \frac{\pi}{2\sin(\tfrac\pi2+\tfrac\pi2\alpha)}
= \tfrac\pi2 \sec \tfrac{\pi \alpha}{2} \ ,
\end{align*}
where we have used the Beta function and Euler's reflection formula. 

Then note that
\[
\cos\lambda + t\sin\lambda + \tfrac{i}2  e^{-i\lambda} \int_0^\lambda e^{(-t+i)s} ds -  \tfrac{i}2  e^{i\lambda} \int_0^\lambda e^{(-t-i)s} ds 
= \frac{e^{-\lambda t}}{t^2+1} \ .
\]
Inserting this identity into the integral \eqref{K int} and reversing the order of integration gives
\begin{align*}
K(\lambda) &= \cos\lambda \int_0^\infty \frac{t^\alpha}{t^2+1}dt + \sin\lambda \int_0^\infty \frac{t^{\alpha+1}}{t^2+1}dt + \tfrac{i}2 e^{-i\lambda}\int_0^\lambda\int_0^\infty t^\alpha e^{-st+is}dt\,ds - \tfrac{i}2 e^{i\lambda}\int_0^\lambda\int_0^\infty t^\alpha e^{-st-is}dt\,ds \\
&= \tfrac\pi2\left(\cos\lambda\sec\tfrac{\pi\alpha}2 - \sin\lambda\csc\tfrac{\pi\alpha}2\right) + \tfrac{i}2\Gamma(\alpha+1) e^{-i\lambda} \int_0^\lambda s^{-\alpha-1}e^{is}ds - \tfrac{i}2\Gamma(\alpha+1) e^{i\lambda} \int_0^\lambda s^{-\alpha-1}e^{-is}ds \ .
\end{align*}
By changing variables to $z=-is$, the first integral becomes
\[
\int_0^\lambda s^{-\alpha-1}e^{is}ds = \int_0^{-i\lambda} (-i z)^{-\alpha-1}e^{-iz} i\,dz
= e^{-\tfrac\pi2\alpha i} \left[\Gamma(-\alpha)-\Gamma(-\alpha,-i\lambda)\right] ,
\]
in terms of the (upper) incomplete Gamma function.

By angle addition formulae,
\[
\tfrac\pi2\left(\cos\lambda\sec\tfrac{\pi\alpha}2 - \sin\lambda\csc\tfrac{\pi\alpha}2\right) 
= \pi \frac{\cos\lambda \sin\frac{\pi\alpha}2 - \sin\lambda \cos\frac{\pi\alpha}2}{2\cos\frac{\pi\alpha}2 \sin\frac{\pi\alpha}2}
= \frac{\pi \sin(\lambda + \tfrac\pi2\alpha)}{\sin \pi\alpha} \ .
\]
By Euler's reflection formula,
\[
\Gamma(\alpha+1)\Gamma(-\alpha) \left(\tfrac{i}2e^{-i\lambda}e^{-i\tfrac\pi2\alpha} - \tfrac{i}2e^{i\lambda}e^{i\tfrac\pi2\alpha}\right) = \pi \csc(-\pi\alpha) \sin\left(\lambda+\tfrac\pi2\alpha\right) = \frac{-\pi \sin(\lambda + \tfrac\pi2\alpha)}{\sin \pi\alpha} \ .
\]
These terms cancel, leaving
\beq
\label{K formula}
K(\lambda) = \Gamma(\alpha+1)\left[\tfrac{i}2 e^{i(\lambda+\tfrac\pi2\alpha)}  \Gamma(-\alpha,i\lambda) -\tfrac{i}2 e^{-i(\lambda+\tfrac\pi2\alpha)} \Gamma(-\alpha,-i\lambda)\right] \ .
\eeq

Next, to evaluate $J(\lambda)$, first suppose that $\lambda\geq0$. The integrand in \eqref{J integral} falls off rapidly enough in the upper half plane that $J(\lambda)$ can be equivalently computed by integrating along the positive imaginary axis.
Setting $q=it$, this gives
\[
J(\lambda) = -i\,e^{i\frac\pi2 \alpha} \int_0^\infty \frac{t^\alpha}{t^2+1}e^{-\lambda t} dt = -i\,e^{i\frac\pi2 \alpha} K(\lambda) \ .
\]
For $\lambda\leq0$, the calculation is very similar, but $J(\lambda)$ equals the integral along the negative imaginary axis, plus the result of integrating \emph{clockwise} around $q=1$, which is $-\pi i\, e^{-i\lambda}$.
Together, these give
\[
J(\lambda) = 
\begin{cases}
 -i\,e^{i\frac\pi2 \alpha} K(\lambda) & \lambda\geq0 \\
 -\pi i\, e^{-i\lambda} +i\,e^{-i\frac\pi2 \alpha} K(-\lambda) & \lambda <0 \ .
 \end{cases}
\]

Now, we can return to the original integral. For $x\geq a$,
\begin{align*}
I(x) 
&= \tfrac{-i\pi}{2}e^{\frac{\pi i}2(-\frac{x}a-1)} + \tfrac{-i\pi}{2}e^{\frac{\pi i}2(-\frac{x}a+1)} + \tfrac1{2}\left(-ie^{i\frac\pi2\alpha}+i e^{-i\frac\pi2\alpha}\right)\left[K(\tfrac\pi2[\tfrac{x}a+1])+K(\tfrac\pi2[\tfrac{x}a-1])\right] \\
&=  \sin\tfrac{\pi \alpha}2 \left[K(\tfrac\pi2[\tfrac{x}a+1])+K(\tfrac\pi2[\tfrac{x}a-1])\right]  \\
&= \tfrac12 \sin\tfrac{\pi \alpha}2\, \Gamma(\alpha+1)\left(e^{\frac{\pi i}{2}(\frac{x}a+\alpha)}\left[\Gamma(-\alpha,\tfrac{\pi i}2[\tfrac{x}a-1])-\Gamma(-\alpha,\tfrac{\pi i}2[\tfrac{x}a+1])\right] + \cc\right) \ .
\end{align*}
For $x\leq-a$, $I(x)=I(-x)$.
For $-a\leq x\leq a$,
\begin{align}
I(x) 
&= \tfrac{-i\pi}{2}e^{\frac{\pi i}2(-\frac{x}a-1)} + \tfrac{-i\pi}{2}e^{\frac{\pi i}2(-\frac{x}a-1)} + \tfrac1{2}\left(-ie^{i\frac\pi2\alpha}+i e^{-i\frac\pi2\alpha}\right)\left[K(\tfrac\pi2[1+\tfrac{x}a])+K(\tfrac\pi2[1-\tfrac{x}a])\right] \nonumber\\
&= -\pi\cos \tfrac{\pi x}{2a} + \sin\tfrac{\pi \alpha}2 \left[K(\tfrac\pi2[1+\tfrac{x}a])+K(\tfrac\pi2[1-\tfrac{x}a])\right] \nonumber\\
&= -\pi \cos \tfrac{\pi x}{2a}  - \tfrac12 \sin\tfrac{\pi \alpha}2\, \Gamma(\alpha+1)\left[e^{\frac{\pi i}{2}(\frac{x}a+\alpha)}\Gamma(-\alpha,\tfrac{\pi i}2[\tfrac{x}a+1]) +  e^{\frac{\pi i}{2}(-\frac{x}a+\alpha)}\Gamma(-\alpha,\tfrac{\pi i}2[-\tfrac{x}a+1])  + \cc\right] \ . \label{I2}
\end{align}

\begin{figure}
\includegraphics{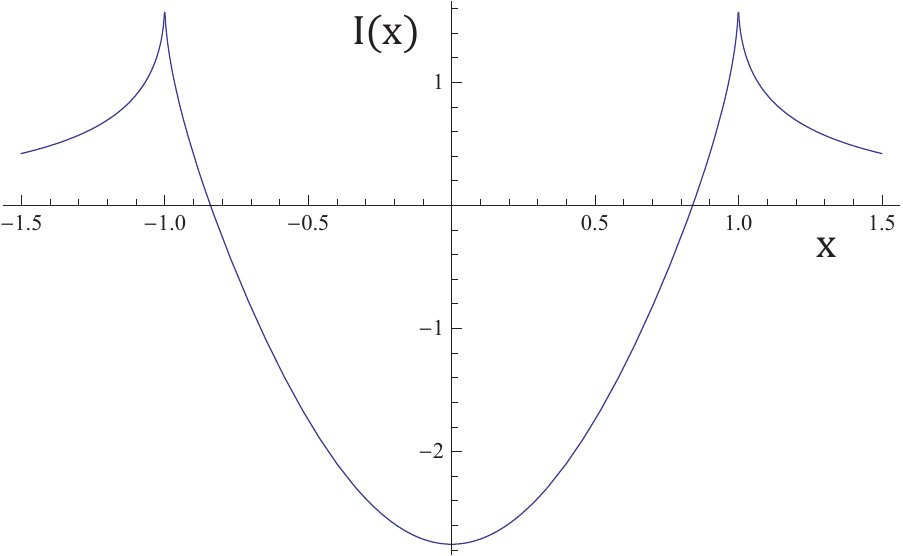}
\caption{The integral $I(x)$ for $\alpha=\frac12$ and $a=1$. Note that this does not vanish at $x=1$.\label{I Plot}}
\end{figure}

If, $\psi_0(x)$, the naive ground state solution to the fractional \Schrodinger\ equation given in Eq.~\eqref{ground state} and Ref.~\cite{Laskin2000} were correct, then \eqref{I2} would be proportional to $\cos \frac{\pi x}{2a}$, but it is not; see Figure~\ref{I Plot}.

\begin{figure}
\includegraphics{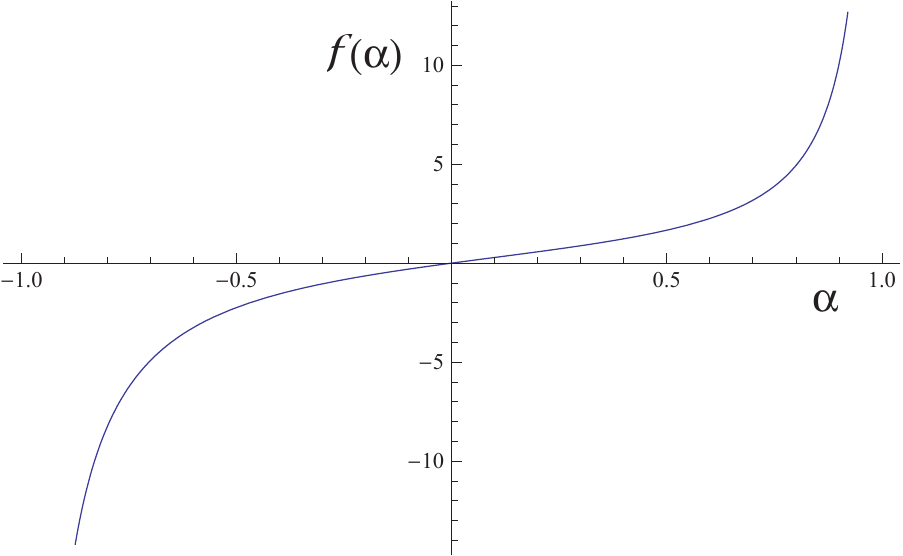}
\caption{The function $f(\alpha)$, which \Bayin\ claims is $0$.\label{f Plot}}
\end{figure}

In particular,
\[
f(\alpha) = \tfrac12 I(a) = \tfrac14 \sin\tfrac{\pi \alpha}2\, \Gamma(\alpha+1) \left(i\, e^{\frac{\pi\alpha i}{2}}\left[\Gamma(-\alpha)-\Gamma(-\alpha,\pi i)\right] + \cc\right) 
\]
is the function that we proved to be monotonically increasing, and which \Bayin\ claims is identically $0$. This function is plotted in Figure~\ref{f Plot}.

Although this computation demonstrates quite explicitly that $\psi_0(x)$ is not a solution of the fractional \Schrodinger\ equation for an infinite square well, it unfortunately does not tell us what the correct solution is. That problem remains open.

\end{document}